\documentclass[aps,prd]{revtex4-2}
\usepackage[utf8]{inputenc}
\usepackage[T1]{fontenc}
\usepackage{lmodern}
\usepackage{amsmath,amssymb,amsfonts,mathtools,amsthm}
\usepackage{geometry}
\usepackage{hyperref}
\usepackage{microtype}
\hypersetup{
  colorlinks=true,
  linkcolor=blue,
  citecolor=blue,
  urlcolor=blue
}

\newcommand{\AdS}{\mathrm{AdS}}
\newcommand{\ESU}{\mathrm{ESU}}
\newcommand{\Mink}{\mathrm{M}}
\newcommand{\dd}{\mathrm{d}}
\newcommand{\OO}{\mathcal{O}}
\newcommand{\Boxop}{\Box}
\newcommand{\sech}{\operatorname{sech}}

\newcommand{\calD}{\mathcal{D}}
\newcommand{\calN}{\mathcal{N}}
\begin{document}
\title{Conformally mapping HKLL to flat space}
 \author{Nirmalya Kajuri}%
\email{nirmalya@iitmandi.ac.in}
 \author{Rhitaparna Pal}
 \email{d21089@students.iitmandi.ac.in}

\affiliation{%
School of Physical Sciences, IIT Mandi,\\ Himachal Pradesh 175005, India}%

\begin{abstract}
We study two conformal images of the HKLL reconstruction problem for a conformally coupled scalar on AdS$_3$. The first mapping is to a Dirichlet problem on one half of the Einstein static universe.  A further conformal transformation maps this half-ESU problem to a massless scalar in a region of $(2+1)$-dimensional Minkowski space bounded by an uniformly accelerated circular Dirichlet mirror. However, the global HKLL problem, the resulting flat-space problem is not a generic local mirror problem with arbitrary boundary data.  Rather, the mirror response data must belong to a restricted spectral class inherited from normalizable global AdS modes.  We formulate and solve this restricted mirror problem intrinsically in flat space. In the process, we find an example of a valid smearing function which itself does not satisfy the equations of motion. We further consider AdS-Rindler wedge reconstruction. By conformal mapping to an appropriate coordinate system in flat space, we find a simple way to see the blow-up of the smearing function.
\end{abstract}
\maketitle
\section{Introduction}

 The AdS/CFT correspondence \cite{Maldacena:1997re,Gubser:1998bc,Witten:1998qj} suggests that all of bulk physics in an asymptotically anti-de Sitter spacetime is encoded in the boundary conformal field theory. HKLL reconstruction \cite{Dobrev:1998md, Bena:1999jv,Hamilton:2005ju,Hamilton:2006az,Hamilton:2006fh,Heemskerk:2012np,Papadodimas:2012aq,Kabat:2011rz,Heemskerk:2012mn, Kabat:2012hp,Leichenauer:2013kaa,Sarkar:2014dma,Sarkar:2014jia,Guica:2014dfa,Roy:2015pga,Kabat:2016rsx,Kabat:2017mun,Kabat:2018pbj,Bhowmick:2018bmn,Foit:2019nsr,Kajuri:2020bvi,Dey:2021vke,Bhattacharjee:2022ehq,Bhattacharjee:2023roq,Kajuri:2020vxf} aims to make this explicit by mapping bulk fields in AdS to a nonlocal smearing of CFT data on the boundary. Solving for the smearing function is essentially a boundary value problem, with the CFT data playing the role of boundary value.

Now, a conformally coupled scalar cannot distinguish between spacetimes in the same
conformal class. This suggests a natural question: what does global AdS reconstruction become after the geometry is conformally flattened? One might expect the HKLL problem to reduce to an ordinary flat-space boundary-value problem. As we will see, the answer is subtler because while conformal covariance preserves the local equation, it need not preserve the global domain on which it is posed. The global conditions selecting the AdS solution must therefore reappear in the flat conformal frame in some other form.  

We study this question for a conformally coupled scalar with  \begin{equation}
\Delta=\frac{3}{2}.
 \end{equation} in global AdS$_3$, using standard quantization. We consider two mappings. The first is a Dirichlet problem on one half of the Einstein static
universe. The second is a scalar in Minkowski space, confined by a
Dirichlet condition to the interior of the timelike hyperboloid
\begin{equation}
  r^2 - t^2 = 1.
  \label{eq:mirror-intro}
\end{equation}
 For the case of the spatially compact Einstein static universe (ESU), global $AdS_3$ is conformal to one hemisphere of it, for all time. The conformal boundary sits at the equator. For the standard quantization (which is what we consider in the paper), the $AdS$ falloff condition becomes an ordinary Dirichlet condition
at the equator while the boundary operator becomes the normal derivative of the
field there. Bulk reconstruction is then the problem of recovering the interior solution from that normal derivative data. Conformal mapping of the HKLL problem was previously studied in \cite{Bhattacharyya:2023czi}, where the Poincare patch was mapped to half-Minkowski space.

For the case of the flat space, \eqref{eq:mirror-intro} shows that the AdS boundary maps to a uniformly accelerated circular mirror. If only the local geometry were relevant, one might expect this to be a generic accelerated-mirror problem. However, conformally flattening the local AdS geometry does not erase its global structure. The latter obstructs such a naive mapping. 

The mismatch can be seen clearly from mapping both AdS and Minkowski spacetimes to ESU.
Global $AdS_3$ occupies a hemisphere for all time, whereas the Minkowski
coordinates cover only a single Penrose diamond of ESU, bounded above and below
by the null surfaces. The global $AdS$ boundary is an
infinite cylinder in $\tau$, and under the conformal map this cylinder splits
into three parts:
\begin{itemize}
  \item the finite segment $-\tfrac{\pi}{2} < \tau < \tfrac{\pi}{2}$, which maps
    to the accelerated mirror $r^2 - t^2 = 1$ inside the diamond;
  \item the two semi-infinite tails $\tau < -\tfrac{\pi}{2}$ and
    $\tau > \tfrac{\pi}{2}$, which are pushed out to the null caps, the past and
    future null boundaries of the diamond.
\end{itemize}

The homogeneous Dirichlet condition on a generic mirror does not by itself select a unique field. One may choose incoming characteristic data on the past null cap; evolution with the mirror condition then determines the resulting mirror response and the outgoing data on the future cap. In the sector obtained from global AdS, however, there is no independent choice of incoming data. Regularity at the centre and the Dirichlet condition at the equator select a global half-ESU mode expansion. Its mode coefficients simultaneously determine the response on the mirror and the characteristic behavior at both null caps.

The intrinsic flat problem is therefore a mirror problem with prescribed characteristic data on the
null caps, the prescription being the one inherited from the global modes.

Once the cap data is fixed, the reconstruction can be posed and solved in flat
space with no reference to AdS.  The resulting smearing function agrees exactly
with the conformal transform of the one on AdS$_3$.

We solve this restricted mirror problem in two stages. We first construct the associated half-ESU solution by expanding the admissible response in global Dirichlet modes and then conformally transform this solution to flat space. We next invert the resulting mode projection to obtain a closed-form kernel on the accelerated mirror. Finally, we independently conformally transform the standard global AdS$_3$ kernel and show that it agrees exactly with the kernel obtained from the restricted mirror problem.

One aspect deserves comment. Solving the equation of motion in both ESU and flat space yields a power-law solution for the smearing function $K_{\mathrm{pow}}(P|b)$ rather than the logarithm that appears in the AdS case. For example, for ESU, we have:
\begin{equation}
\left(\Box_{\mathrm{ESU}}+\frac14\right)
K_{\mathrm{pow}}(P|b)=0,
\end{equation}
where the homogenous wave operator on the LHS is the one obtained via conformal transport from AdS$_3$.
There is no logarithmic solution to this homogenous wave equation. At first sight, this looks like an inconsistency.

The resolution lies in the ambiguity of the smearing function. It is a well known aspect of HKLL reconstruction that one may add to the smearing function any function that integrates to zero against them without changing the reconstructed field. The power-law solution turns out to be exactly of this type — it integrates to zero against every admissible mode:
\begin{equation}
\int db\,
K_{\mathrm{pow}}(P|b)\,O_D(b)=0,
\qquad
O_D\in\mathcal D_{\mathrm{AdS}}.
\end{equation}
A nonzero reconstruction is obtained only after introducing a logarithmic  kernel $K_{\log}(P|b)$. 

This logarithmic term does not solve the equation of motion by itself, but leaves a residual proportional to the power-law kernel:
\begin{equation}
\left(\Box_{\mathrm{ESU}}+\frac14\right)
K_{\log}(P|b)
   =-z^{-2}K_{\mathrm{pow}}(P|b).
\end{equation}
But as the power-law kernel vanishes when smeared against admissible boundary data, the reconstructed field obtained by convolution of the logarithmic term with admissible boundary data does satisfy the required bulk equation:
\begin{align}
\left(\Box_{\mathrm{ESU}}+\frac14\right)
\Phi_{\mathrm{ESU}}(P)
&=
\int db\,
\left(\Box_{\mathrm{ESU}}+\frac14\right)
K_{\log}(P|b)\,O_D(b) \\
&=
-z^{-2}\int db\,
K_{\mathrm{pow}}(P|b)\,O_D(b) \\
&=0.
\end{align}

We highlight this because in HKLL literature, it is usually claimed that the smearing function solve the equations of motion (indeed, solving the EoM is how smearing functions are typically constructed). However, as this example underlines, smearing functions are only required to solve the EoM up to the known ambiguities in smearing function. 

In this paper, we also apply the flat-space conformal map to the reconstruction problem for a boundary subregion. In AdS--Rindler reconstruction, one attempts to recover a bulk field in the causal wedge of a boundary ball from data on its boundary diamond. Although the field is fixed mode by mode, the ordinary real-boundary smearing integral diverges because modes with large transverse momentum are exponentially amplified inside the wedge \cite{Bousso:2012mh,Leichenauer:2013kaa,Rey:2014dpa,Morrison:2014jha,Sugishita:2022ldv}. We show that the complete AdS--Rindler wedge and its entire boundary diamond map into the Minkowski patch of the ESU, where they become the interior to the uniformly accelerated mirror and the corresponding partial mirror surface. The flat description provides a simpler setting in which to understand the obstruction. In polar Rindler coordinates the mirror lies at constant radius, and the response-normalized radial modes become elementary power laws, making the exponential instability transparent without using the usual AdS--Rindler hypergeometric modes. We use a \(\mathbb R\times H^2\) frame to connects this direct flat-space calculation back to the standard AdS--Rindler result.

The remainder of the paper is organized as follows. Section~II reviews global AdS$_3$ HKLL reconstruction for the conformally coupled scalar. Section~III reformulates the problem on the half-ESU and derives the corresponding smearing kernel. Section~IV introduces the conformal map to Minkowski space and defines the admissible accelerated-mirror responses. Section~V solves the restricted mirror problem and obtains its closed-form kernel. Section~VI shows that this kernel agrees with the conformal image of the global AdS$_3$ HKLL kernel. Section~VII discusses the AdS-Rindler map. Section~VIII summarizes the results and discusses possible extensions.

\section{Standard AdS$_3$ HKLL for the conformally coupled scalar}
This section provides a brief review of HKLL construction in AdS$_3$.
We use unit AdS radius and mostly-minus signature.  Global AdS$_3$ has metric
\begin{equation}
  \dd s^2_{\AdS}
  =
  \frac{1}{\cos^2\rho}
  \left(
    \dd\tau^2-\dd\rho^2-\sin^2\rho\,\dd\theta^2
  \right),
  \qquad
  0\leq \rho<\frac{\pi}{2},
  \qquad
  \theta\sim\theta+2\pi .
\end{equation}
We define
\begin{equation}
  z=\cos\rho .
\end{equation}

A conformally coupled massless scalar in three spacetime dimensions corresponds in AdS units to
\begin{equation}
  m^2_{\AdS}=-\frac{3}{4} .
\end{equation}
The two possible conformal dimensions are
\begin{equation}
  \Delta_{\pm}=1\pm\sqrt{1+m^2_{\AdS}}
  =\frac{1}{2},\frac{3}{2} .
\end{equation}
In this paper we use only the standard quantization:
\begin{equation}
  \Delta=\Delta_+=\frac{3}{2} .
\end{equation}
The boundary operator is defined by
\begin{equation}
  \OO(\tau,\theta)
  =
  \lim_{\rho\to\pi/2}z^{-3/2}\Phi_{\AdS}(\tau,\rho,\theta) .
\end{equation}

The regular normalizable modes are
\begin{equation}
  \Phi_{n j}(\tau,\rho,\theta)
  =
  e^{-i\omega_{n j}\tau}
  e^{i j\theta}
  z^{3/2}
  (\sin\rho)^{|j|}
  P_n^{(1/2,|j|)}(-\cos 2\rho),
\end{equation}
with
\begin{equation}
  \omega_{n j}=\frac{3}{2}+|j|+2n,
  \qquad
  n=0,1,2,\ldots,
  \qquad
  j\in\mathbb Z .
\end{equation}
Thus
\begin{equation}
  \Phi_{\AdS}
  =
  \sum_{n,j}
  \left(
    a_{n j}\Phi_{n j}+a_{n j}^{\dagger}\Phi_{n j}^{*}
  \right).
\end{equation}
Near the boundary,
\begin{equation}
  \Phi_{n j}
  \sim
  z^{3/2}
  e^{-i\omega_{n j}\tau}
  e^{i j\theta}
  P_n^{(1/2,|j|)}(1),
\end{equation}
so the normalizable-mode coefficients are encoded in \(\OO(\tau,\theta)\).

The global HKLL reconstruction takes the form
\begin{equation}
  \Phi_{\AdS}(P)
  =
  \int \dd\tau'\,\dd\theta'\,
  K_{\AdS}(P|\tau',\theta')\,
  \OO(\tau',\theta') .
\end{equation}
For the odd-dimensional global smearing representative appropriate to AdS$_3$, the kernel may be written as
\begin{equation}
  K_{\AdS}(P|b)
  =
  \calN\,
  \Theta(Q)
  \left(\frac{Q}{z}\right)^{-1/2}
  \log\left(\frac{Q}{z}\right),
  \label{eq:AdSKernel}
\end{equation}
where \(\calN\) is an overall normalization fixed by the convention for \(\OO\), and
\begin{equation}
  b=(\tau',\theta'),
\end{equation}
\begin{equation}
  Q(P|b)
  =
  \cos(\tau-\tau')
  -
  \sin\rho\cos(\theta-\theta') .
  \label{eq:Qdef}
\end{equation}
Equivalently,
\begin{equation}
  K_{\AdS}(P|b)
  =
  \calN\,
  \Theta(Q)
  \frac{\sqrt z}{\sqrt Q}
  \log\left(\frac{Q}{z}\right).
\end{equation}
The support condition \(Q>0\) is the usual spacelike HKLL support condition in this representative.  The smearing representative is not unique: one may add terms that integrate to zero against the allowed global boundary modes.  This ambiguity is inherited by the conformally related kernels below.

\section{The half-ESU Dirichlet problem}

In this section we map the AdS$_3$ reconstruction problem to the half-ESU and
solve it there, using only the ESU equation of motion and the Dirichlet
condition at the equator.  The AdS smearing function is not used; it reappears
at the end as a consistency check.

\subsection{Reduction to a Dirichlet problem}

The Einstein static universe metric is
\begin{equation}
  \dd s^2_{\ESU}
  =
  \dd\tau^2-\dd\rho^2-\sin^2\rho\,\dd\theta^2 ,
\end{equation}
conformally related to global AdS$_3$ by
\begin{equation}
  \dd s^2_{\ESU}=z^2\dd s^2_{\AdS},
  \qquad
  z=\cos\rho .
\end{equation}
For a conformally coupled scalar in three spacetime dimensions
\(\Phi_{\AdS}=z^{1/2}\Phi_{\ESU}\), and the equation of motion becomes
\begin{equation}
  \left(\Boxop_{\ESU}+\frac14\right)\Phi_{\ESU}=0 ,
  \label{eq:ESUeom}
\end{equation}
the conformally coupled wave equation on \(\mathbb{R}\times S^2\).  The AdS region
\(0\leq\rho<\pi/2\) is the half-ESU and the AdS conformal boundary is the equator
\(\rho=\pi/2\).  In the standard quantization \(\Phi_{\AdS}\sim z^{3/2}\OO\), so
\begin{equation}
  \Phi_{\ESU}=z^{-1/2}\Phi_{\AdS}\sim z\,\OO(\tau,\theta),
\end{equation}
and \(\Phi_{\ESU}\) obeys a Dirichlet condition at the equator,
\begin{equation}
  \Phi_{\ESU}\big|_{\rho=\pi/2}=0,
\end{equation}
with the AdS boundary operator identified with the Dirichlet response
\begin{equation}
  \OO_D(\tau,\theta)
  =
  -\partial_{\rho}\Phi_{\ESU}\big|_{\rho=\pi/2}
  =
  \OO(\tau,\theta) .
  \label{eq:ODdef}
\end{equation}
Reconstruction is the inverse problem of recovering the interior field from this
response,
\begin{equation}
  \Phi_{\ESU}(P)
  =
  \int \dd\tau'\,\dd\theta'\,
  K_D^{\ESU}(P|b)\,\OO_D(b) ,
  \label{eq:ESUrecon}
\end{equation}
and it is \(K_D^{\ESU}\) that we now determine from \eqref{eq:ESUeom} and
\eqref{eq:ODdef}.

\subsection{Spectrum}

Separating variables, \(\Phi_{\ESU}=e^{-i\omega\tau}e^{ij\theta}R(\rho)\),
\eqref{eq:ESUeom} becomes the associated Legendre equation
\begin{equation}
  R''+\cot\rho\,R'-\frac{j^2}{\sin^2\rho}\,R
  +\left(\omega^2-\tfrac14\right)R=0,
\end{equation}
whose solution regular at \(\rho=0\) is \(R=P_\ell^{|j|}(\cos\rho)\) with
\(\omega=\ell+\tfrac12\), \(\ell=0,1,2,\ldots\).  The Dirichlet condition selects
the modes vanishing at the equator, and since
\begin{equation}
  P_{\ell}^{|j|}(0)=0
  \quad\Longleftrightarrow\quad
  \ell+|j|\ \text{odd},
\end{equation}
the Dirichlet sector is \(\ell=|j|+2n+1\), i.e.
\begin{equation}
  \omega_{n j}=\ell+\tfrac12=|j|+2n+\tfrac32,
  \qquad n=0,1,2,\ldots,
  \label{eq:ESUfreq}
\end{equation}
the standard AdS$_3$ frequencies for \(\Delta=3/2\), obtained here as a parity
selection rule at the equator.

\subsection{The smearing function from the equation of motion}

Because \(\Phi_{\ESU}(P)\) solves \eqref{eq:ESUeom} while \(\OO_D(b)\) is
independent of \(P\), the kernel in \eqref{eq:ESUrecon} must be annihilated by
\((\Boxop_{\ESU}+\tfrac14)\) acting on \(P\), when integrated against admissible
data.  Invariance under \(\tau\)-translation and \(\theta\)-rotation restricts
its dependence to \(Q\) of \eqref{eq:Qdef} and the radial scalar \(z=\cos\rho\).
A direct computation gives, for any power,
\begin{equation}
  \left(\Boxop_{\ESU}+\tfrac14\right)Q^{s}
  =
  -s\big(s+\tfrac12\big)\,\bar Q\,Q^{\,s-1}
  +\tfrac12\big(s+\tfrac12\big)\,Q^{s},
  \qquad
  \bar Q=\cos(\tau-\tau')+\sin\rho\cos(\theta-\theta') .
  \label{eq:poweraction}
\end{equation}
Both coefficients vanish only at \(s=-\tfrac12\).  The equation of motion
therefore selects, among spacelike-supported invariant power laws, the unique
representative \(\Theta(Q)\,Q^{-1/2}\), the step function imposing \(Q>0\).

This representative solves the equation but reconstructs nothing.  Its projection
on a mode is, by \eqref{eq:poweraction} and invariance,
\begin{equation}
  I_{\omega j}(\tau,\rho,\theta)
  =\int \dd\tau'\dd\theta'\,\Theta(Q)\,Q^{-1/2}
  e^{-i\omega\tau'}e^{ij\theta'}
  =e^{-i\omega\tau}e^{ij\theta}\,{\cal I}_{\omega j}(\rho),
\end{equation}
and away from \(Q=0\) the integrand is annihilated by
\((\Boxop_{\ESU}+\tfrac14)\), so \({\cal I}_{\omega j}\) obeys the Legendre
equation of degree \(\omega-\tfrac12\).  Regularity at \(\rho=0\) fixes
\({\cal I}_{\omega j}={\cal N}_{\omega j}\,P^{|j|}_{\omega-1/2}(\cos\rho)\), with
\begin{equation}
  {\cal N}_{\omega j}
  \propto
  \frac{1}{\Gamma\!\left(\frac{|j|+3/2+\omega}{2}\right)
           \Gamma\!\left(\frac{|j|+3/2-\omega}{2}\right)} .
\end{equation}
On the Dirichlet spectrum \eqref{eq:ESUfreq},
\(\tfrac12(|j|+\tfrac32-\omega)=-n\), so \(\Gamma(-n)\) has a pole and
\begin{equation}
  \int \dd\tau'\dd\theta'\,\Theta(Q)\,Q^{-1/2}
  e^{-i(\ell+1/2)\tau'}e^{ij\theta'}=0,
  \qquad \ell=|j|+2n+1 .
  \label{eq:spectralzero}
\end{equation}
The same zero is visible at \(j=0\): at \(\rho=\tau=0\), \(Q=\cos\tau'\), and the
beta-function identity
\begin{equation}
  \int_{-\pi/2}^{\pi/2}\dd\tau\,(\cos\tau)^{\nu-1}e^{-i\omega\tau}
  =\frac{\pi\,\Gamma(\nu)}
        {2^{\nu-1}\Gamma\!\left(\frac{\nu+1+\omega}{2}\right)
                  \Gamma\!\left(\frac{\nu+1-\omega}{2}\right)}
  \label{eq:betaidentity}
\end{equation}
at \(\nu=\tfrac12\) has \(\Gamma\!\left(\tfrac{3/2-\omega}{2}\right)=\Gamma(-n)\)
on \(\omega=2n+\tfrac32\).

A logarithmic representative is therefore required.  With
\(\Theta(Q)\,Q^{-1/2}\log\!\big(Q/w(\rho)\big)\) and \(G=\log w\),
\begin{equation}
  \left(\Boxop_{\ESU}+\tfrac14\right)\!\left[Q^{-1/2}\log\frac{Q}{w}\right]
  =\big(1+G'\cot\rho\big)\cos(\tau-\tau')\,Q^{-3/2}+G''\,Q^{-1/2} .
  \label{eq:logaction}
\end{equation}
The \(Q^{-3/2}\) term is more singular than \eqref{eq:spectralzero} and does not
lie in the annihilated subspace; requiring it to vanish fixes \(G'=-\tan\rho\),
i.e.
\begin{equation}
  w(\rho)=\cos\rho=z,
\end{equation}
uniquely up to a multiplicative constant, which shifts the kernel by a multiple
of \(\Theta(Q)Q^{-1/2}\) and is immaterial by \eqref{eq:spectralzero}.  The
residual source,
\begin{equation}
  \left(\Boxop_{\ESU}+\tfrac14\right)\!\left[Q^{-1/2}\log\frac{Q}{z}\right]
  =-\,z^{-2}\,Q^{-1/2},
\end{equation}
is proportional to the spectral-zero representative and is therefore annihilated
by admissible data: \(\Phi_{\ESU}=\int K_D^{\ESU}\OO_D\) solves \eqref{eq:ESUeom}
on \(\calD_{\AdS}\).

The normalization is fixed by the Dirichlet-response condition.  With
\begin{equation}
  R_{\ell j}(\rho)=\frac{P_\ell^{|j|}(\cos\rho)}{D_{\ell j}},
  \qquad
  D_{\ell j}=\left.\frac{\dd}{\dd x}P_\ell^{|j|}(x)\right|_{x=0},
  \qquad
  -\partial_\rho R_{\ell j}\big|_{\rho=\pi/2}=1,
  \label{eq:Rdef}
\end{equation}
an elementary beta-function evaluation of the log-projection gives the half-ESU
identity
\begin{equation}
  \boxed{
  \int \dd\tau'\dd\theta'\,
  \Theta(Q)\,\frac{1}{\sqrt Q}\log\frac{Q}{z}\,
  e^{-i(\ell+1/2)\tau'}e^{ij\theta'}
  =
  {\cal C}\,e^{-i(\ell+1/2)\tau}e^{ij\theta}\,R_{\ell j}(\rho),
  \qquad
  {\cal C}=2\sqrt2\,\pi^2 .
  }
  \label{eq:ESUmaster}
\end{equation}
Hence
\begin{equation}
  K_D^{\ESU}(P|b)
  =
  \calN\,\Theta(Q)\,\frac{1}{\sqrt Q}\log\!\left(\frac{Q}{z}\right),
  \label{eq:ESUKernel}
\end{equation}
with \(\calN\) fixed by the convention for \(\OO_D\).  Since
\(\Phi_{\AdS}=z^{1/2}\Phi_{\ESU}\), this predicts
\(K_{\AdS}=z^{1/2}K_D^{\ESU}
=\calN\,\Theta(Q)(Q/z)^{-1/2}\log(Q/z)\), which is the standard kernel
\eqref{eq:AdSKernel}.

\section{The flat space mirror problem}

In this section, we map to the flat space mirror problem. First, we define the map then proceed to solve the problem.

\subsection{The conformal map}

Let
\begin{equation}
  \dd s^2_{\Mink}
  =
  \dd t^2-\dd r^2-r^2\dd\theta^2
\end{equation}
be the $(2+1)$-dimensional Minkowski metric.  Introduce
\begin{equation}
  u=t-r,
  \qquad
  v=t+r,
\end{equation}
and compact coordinates
\begin{equation}
  \tau=\arctan v+\arctan u,
  \qquad
  \rho=\arctan v-\arctan u .
  \label{eq:compactcoords}
\end{equation}
Equivalently,
\begin{equation}
  t=\frac{\sin\tau}{\Omega},
  \qquad
  r=\frac{\sin\rho}{\Omega},
  \qquad
  \Omega=\cos\tau+\cos\rho .
\end{equation}
Then
\begin{equation}
  \dd s^2_{\ESU}=\Omega^2\dd s^2_{\Mink} .
\end{equation}
Since \(\dd s^2_{\ESU}=z^2\dd s^2_{\AdS}\), we also have
\begin{equation}
  \dd s^2_{\Mink}
  =
  \left(\frac{z}{\Omega}\right)^2\dd s^2_{\AdS} .
\end{equation}
Define
\begin{equation}
  \Lambda(x)=\frac{z}{\Omega} .
\end{equation}
In flat coordinates,
\begin{equation}
  \Lambda(x)=\frac{1-r^2+t^2}{2} .
  \label{eq:Lambdadef}
\end{equation}
For a conformally coupled scalar in three dimensions,
\begin{equation}
  \Phi_{\Mink}
  =
  \Lambda^{-1/2}\Phi_{\AdS}
  =
  \Omega^{1/2}\Phi_{\ESU} .
  \label{eq:scalarWeylFlat}
\end{equation}

The AdS boundary \(z=0\), equivalently \(\rho=\pi/2\), maps to
\begin{equation}
  r^2-t^2=1 .
\end{equation}
The AdS bulk maps to
\begin{equation}
  D=\{(t,r,\theta): r^2-t^2<1\} .
\end{equation}
Thus the conformal image of the standard AdS Dirichlet problem is a massless scalar in \(D\) with a timelike Dirichlet mirror
\begin{equation}
  \Sigma:\quad r^2-t^2=1 .
\end{equation}

Parametrize the mirror by the rapidity \(\eta\) as
\begin{equation}
  B^{\mu}(\eta,\theta)
  =
  (\sinh\eta,\cosh\eta\cos\theta,\cosh\eta\sin\theta) .
  \label{eq:mirrorparam}
\end{equation}
On the mirror,
\begin{equation}
  \tau=\tau(\eta)
  =
  2\arctan(e^{\eta})-\frac{\pi}{2},
\end{equation}
and
\begin{equation}
  \frac{\dd\tau}{\dd\eta}=\sech\eta,
  \qquad
  \cos\tau(\eta)=\sech\eta .
  \label{eq:taueta}
\end{equation}

Approach the mirror from the interior by
\begin{equation}
  x^{\mu}=\lambda B^{\mu}(\eta,\theta),
  \qquad
  \lambda\to 1^{-} .
\end{equation}
Since \(r^2-t^2=\lambda^2\), we have
\begin{equation}
  \Lambda(x)=\frac{1-\lambda^2}{2}\sim 1-\lambda .
\end{equation}
The flat field obeys the Dirichlet condition
\begin{equation}
  \Phi_{\Mink}\big|_{\Sigma}=0 .
\end{equation}
The natural mirror response is
\begin{equation}
  \OO_{\Mink}(\eta,\theta)
  =
  \lim_{\lambda\to1^{-}}
  \frac{\Phi_{\Mink}(\lambda B(\eta,\theta))}{1-\lambda} .
  \label{eq:OMdef}
\end{equation}
Equivalently,
\begin{equation}
  \OO_{\Mink}(\eta,\theta)
  =
  -\partial_{\lambda}\Phi_{\Mink}(\lambda B(\eta,\theta))\big|_{\lambda=1} .
\end{equation}
Using \eqref{eq:scalarWeylFlat}, \eqref{eq:ODdef}, and \eqref{eq:taueta}, we find
\begin{equation}
  \Phi_{\Mink}(\lambda B(\eta,\theta))
  \sim
  (1-\lambda)\sech^{3/2}\eta\,
  \OO_D(\tau(\eta),\theta) .
\end{equation}
Therefore
\begin{equation}
  \boxed{
  \OO_{\Mink}(\eta,\theta)
  =
  \sech^{3/2}\eta\,
  \OO_D(\tau(\eta),\theta)
  }
  \label{eq:OMOD}
\end{equation}
Since \(\OO_D=\OO\), this is also
\begin{equation}
  \OO_{\Mink}(\eta,\theta)
  =
  \sech^{3/2}\eta\,
  \OO(\tau(\eta),\theta) .
\end{equation}
This gives the translation of the AdS boundary data into flat mirror data.
 
\subsection{The restricted flat mirror problem}

We now formulate the flat problem without directly referring to AdS.  Let
\begin{equation}
  D=\{r^2-t^2<1\}
\end{equation}
be the interior of the accelerated circular mirror.  We seek massless solutions
\begin{equation}
  \Boxop_{\Mink}\Phi_{\Mink}=0
  \qquad\text{in }D,
\end{equation}
obeying
\begin{equation}
  \Phi_{\Mink}\big|_{\Sigma}=0,
  \qquad
  \Sigma: r^2-t^2=1 .
\end{equation}
The boundary response is
\begin{equation}
  F(\eta,\theta)
  =
  -\partial_{\lambda}\Phi_{\Mink}(\lambda B(\eta,\theta))\big|_{\lambda=1} .
\end{equation}

The region \(D\) is bounded by the mirror on the timelike side and by the two
null caps on the past and future sides.  The mirror problem is well-posed only
once data is specified on both: the Dirichlet condition on \(\Sigma\) and
characteristic data on the null caps.  For a generic mirror problem, the cap
data is free.  The global AdS problem fixes it: the mode expansion, determined
by regularity at the centre and the Dirichlet condition at the equator,
propagates specific nonzero data onto the caps.  A compatible response \(F\) is
one that arises from such a global configuration.

Concretely, define
\begin{equation}
  f(\tau,\theta)
  =
  \cos^{-3/2}\tau\,
  F(\eta(\tau),\theta),
  \qquad
  -\frac{\pi}{2}<\tau<\frac{\pi}{2} .
  \label{eq:fdata}
\end{equation}
We say that \(F\) belongs to the admissible data space \(\calD_{\AdS}\) if
\(f(\tau,\theta)\) is the restriction to \(-\pi/2<\tau<\pi/2\) of a global
half-ESU Dirichlet response, that is, a function of the form
\begin{equation}
  f_{\rm gl}(\tau,\theta)
  =
  \sum_{j\in\mathbb Z}\sum_{n=0}^{\infty}
  \left[
    c_{n j}
    e^{-i(|j|+2n+3/2)\tau}
    e^{i j\theta}
    +
    c_{n j}^{*}
    e^{i(|j|+2n+3/2)\tau}
    e^{-i j\theta}
  \right].
  \label{eq:restrictedSpectrum}
\end{equation}
Equivalently,
\begin{equation}
  F(\eta,\theta)
  =
  \sech^{3/2}\eta\,
  f_{\rm gl}(\tau(\eta),\theta) .
\end{equation}

The discrete spectrum
\begin{equation}
  \omega_{n j}=|j|+2n+\frac32
\end{equation}
is not an additional input.  It arises from the compactification: the global
expansion \eqref{eq:restrictedSpectrum} lives on the full ESU cylinder, whose
spatial sections are compact, and the frequencies are the eigenvalues of the
Laplacian on the hemisphere with Dirichlet boundary conditions.  In the flat
problem, the same information is encoded not as a spectral condition but as the
specific characteristic data that the global modes deposit on the null caps.
The coefficients \(c_{n j}\) that fix the mirror response also fix the field on
the caps---no independent data enters there.  The reconstruction problem is to
recover the bulk field from the admissible response \(F\) alone, with the
understanding that the cap data is already determined by it.

 \section{Solution of the restricted mirror problem}

We first solve the restricted problem as a mode sum, then package the solution
as a closed-form kernel by reducing the flat projection to the half-ESU identity
\eqref{eq:ESUmaster}.

\subsection{Mode-sum solution}

Given \(F\in\calD_{\AdS}\) with coefficients \(c_{n j}\) as in
\eqref{eq:restrictedSpectrum}, set \(\ell=|j|+2n+1\), \(\omega_{n j}=\ell+\tfrac12\).
The half-ESU solution with Dirichlet response \(f_{\rm gl}\) is
\begin{equation}
  \Phi_{\ESU}(\tau,\rho,\theta)
  =
  \sum_{j\in\mathbb Z}\sum_{n=0}^{\infty}
  \left[
    c_{n j}e^{-i\omega_{n j}\tau}e^{i j\theta}
    +
    c_{n j}^{*}e^{i\omega_{n j}\tau}e^{-i j\theta}
  \right]
  R_{\ell j}(\rho),
  \label{eq:ESUmodesolution}
\end{equation}
with \(R_{\ell j}\) as in \eqref{eq:Rdef}.  It satisfies
\((\Boxop_{\ESU}+\tfrac14)\Phi_{\ESU}=0\),
\(\Phi_{\ESU}|_{\rho=\pi/2}=0\), and
\(-\partial_{\rho}\Phi_{\ESU}|_{\rho=\pi/2}=f_{\rm gl}\).  The corresponding
flat-space solution is
\begin{equation}
  \boxed{
  \Phi_{\Mink}(t,r,\theta)
  =
  \Omega^{1/2}(\tau,\rho)\,
  \Phi_{\ESU}(\tau,\rho,\theta)
  }
  \label{eq:flatmodesolution}
\end{equation}
with \((\tau,\rho)\) obtained from \((t,r)\) by \eqref{eq:compactcoords}.  This is
an intrinsic solution of the restricted flat mirror problem: the equation solved
is the flat wave equation with a Dirichlet accelerated mirror, and the nonlocal
input is the restricted spectral class of allowed responses.

\subsection{Closed-form kernel}

The mode sum does not yet give a kernel.  We invert the spectral transform by
reducing the flat projection to \eqref{eq:ESUmaster}.  Let
\begin{equation}
  \Psi_{\ell j}(\eta,\theta)
  =
  \sech^{3/2}\eta\,
  e^{-i(\ell+1/2)\tau(\eta)}e^{ij\theta},
  \qquad
  \ell=|j|+2n+1 ,
\end{equation}
with bulk solution \(\Phi^M_{\ell j}(x)=\Omega^{1/2}(x)e^{-i(\ell+1/2)\tau}
e^{ij\theta}R_{\ell j}(\rho)\).  A flat kernel must obey
\(\int \dd\eta'\dd\theta'\,K_D^M(x|\eta',\theta')\,\Psi_{\ell j}
=\Phi^M_{\ell j}(x)\).  Define
\begin{equation}
  \sigma_b(x|\eta',\theta')
  =
  \frac12\left(1+r^2-t^2+2t\sinh\eta'-2r\cosh\eta'\cos(\theta-\theta')\right),
  \label{eq:sigmab}
\end{equation}
and the unnormalised kernel
\begin{equation}
  \widehat K_D^M(x|\eta',\theta')
  =
  \cosh\eta'\,\Theta(\sigma_b)\,\frac{1}{\sqrt{\sigma_b}}
  \log\!\left[\frac{\sigma_b}{\Lambda(x)\cosh\eta'}\right] .
\end{equation}
Using \(\dd\eta'=\cosh\eta'\,\dd\tau'\),
\(\Psi_{\ell j}=\sech^{3/2}\eta'\,e^{-i(\ell+1/2)\tau'}e^{ij\theta'}\), and
\begin{equation}
  Q=\frac{\Omega(x)\,\sigma_b}{\cosh\eta'},
  \qquad
  \frac{Q}{z}=\frac{\sigma_b}{\Lambda(x)\cosh\eta'},
  \label{eq:Qzsigmab}
\end{equation}
the projection becomes
\begin{align}
  \int \dd\eta'\dd\theta'\,
  \widehat K_D^M\,\Psi_{\ell j}
  &=
  \Omega^{1/2}(x)\int \dd\tau'\dd\theta'\,
  \Theta(Q)\,\frac{1}{\sqrt Q}\log\!\left(\frac{Q}{z}\right)
  e^{-i(\ell+1/2)\tau'}e^{ij\theta'}
  \nonumber\\
  &=
  {\cal C}\,\Omega^{1/2}(x)\,e^{-i(\ell+1/2)\tau}e^{ij\theta}R_{\ell j}(\rho)
  =
  {\cal C}\,\Phi^M_{\ell j}(x),
\end{align}
by \eqref{eq:ESUmaster}.  Hence the properly normalised kernel is
\begin{equation}
  \boxed{
  K_D^{\Mink}(x|\eta',\theta')
  =
  N\,\cosh\eta'\,\Theta(\sigma_b)\,\frac{1}{\sqrt{\sigma_b}}
  \log\!\left[\frac{\sigma_b}{\Lambda(x)\cosh\eta'}\right]
  }
  \qquad
  N=\frac1{\cal C}=\frac{1}{2\sqrt2\,\pi^2}.
  \label{eq:flatkernel}
\end{equation}
Using \(\Lambda=(1-r^2+t^2)/2\),
\begin{equation}
  K_D^{\Mink}(x|\eta',\theta')
  =
  N\,\cosh\eta'\,\Theta(\sigma_b)\,\frac{1}{\sqrt{\sigma_b}}
  \log\!\left[\frac{2\sigma_b}{(1-r^2+t^2)\cosh\eta'}\right] .
\end{equation}
The reconstruction is
\begin{equation}
  \Phi_{\Mink}(t,r,\theta)
  =
  \int_{-\infty}^{\infty}\dd\eta'\int_0^{2\pi}\dd\theta'\,
  K_D^{\Mink}(t,r,\theta|\eta',\theta')\,\OO_{\Mink}(\eta',\theta'),
  \qquad \OO_{\Mink}\in\calD_{\AdS} .
\end{equation}
This used only the restricted flat modes and their half-ESU form; the AdS kernel
was not assumed.

\section{Matching with the conformal image of AdS HKLL}

We now compare the flat mirror kernel obtained by the restricted mode projection with the conformal transform of the AdS HKLL kernel.

The AdS reconstruction is
\begin{equation}
  \Phi_{\AdS}(P)
  =
  \int\dd\tau'\,\dd\theta'\,
  K_{\AdS}(P|\tau',\theta')\,
  \OO(\tau',\theta') .
\end{equation}
Using
\begin{equation}
  \Phi_{\Mink}(x)=\Lambda(x)^{-1/2}\Phi_{\AdS}(P(x)),
\end{equation}
\begin{equation}
  \dd\tau'=\sech\eta'\,\dd\eta',
\end{equation}
and
\begin{equation}
  \OO(\tau(\eta'),\theta')
  =
  \cosh^{3/2}\eta'\,
  \OO_{\Mink}(\eta',\theta'),
\end{equation}
we obtain
\begin{equation}
  \Phi_{\Mink}(x)
  =
  \int \dd\eta'\,\dd\theta'\,
  \Lambda(x)^{-1/2}
  \sqrt{\cosh\eta'}\,
  K_{\AdS}(P(x)|\tau(\eta'),\theta')\,
  \OO_{\Mink}(\eta',\theta') .
\end{equation}
Hence the conformally transformed AdS kernel is
\begin{equation}
  \boxed{
  K^{\Mink}_{\rm conf}(x|\eta',\theta')
  =
  \Lambda(x)^{-1/2}
  \sqrt{\cosh\eta'}\,
  K_{\AdS}(P(x)|\tau(\eta'),\theta')
  }
  \label{eq:confkernel}
\end{equation}
Using \eqref{eq:AdSKernel} and \eqref{eq:Qzsigmab}, we find
\begin{align}
  K^{\Mink}_{\rm conf}
  & =
  \calN\,
  \Lambda^{-1/2}
  \sqrt{\cosh\eta'}\,
  \Theta(\sigma_b)
  \left(
    \frac{\sigma_b}{\Lambda\cosh\eta'}
  \right)^{-1/2}
  \log\left[
    \frac{\sigma_b}{\Lambda\cosh\eta'}
  \right] \\
  & =
  \calN\,
  \cosh\eta'\,
  \Theta(\sigma_b)
  \frac{1}{\sqrt{\sigma_b}}
  \log\left[
    \frac{\sigma_b}{\Lambda(x)\cosh\eta'}
  \right] .
\end{align}
This is exactly the kernel \eqref{eq:flatkernel} found by solving the restricted flat mirror problem:
\begin{equation}
  \boxed{
  K^{\Mink}_{\rm conf}=K_D^{\Mink} .
  }
\end{equation}
Thus the conformal transform of standard AdS$_3$ HKLL agrees with the independent restricted mirror reconstruction.
\section{The AdS--Rindler wedge as a partial mirror problem}
\label{sec:rindler}

So far, we have considered reconstruction for the complete AdS spacetime. We now consider the case of causal wedge reconstruction and map that problem to flat space. It is well known that for causal wedge reconstruction, the smearing function diverges. This should also hold for the flat space problem.

More specifically, we restrict the reconstruction data to the
domain of dependence of a boundary interval. The associated causal wedge is
isometric to an AdS--Rindler wedge. Under the conformal map used in this paper,
the complete open wedge lies inside the Minkowski patch of the ESU; no part of
the wedge requires data from outside that patch.

The reconstruction data consist of the homogeneous Dirichlet condition
together with the normal-derivative response on the boundary diamond. These
two pieces of boundary information uniquely determine every radial mode. The
question is if
the resulting boundary-to-bulk map is sufficiently well behaved to be
represented by an ordinary smearing function.

We proceed in three steps. First, we show that the complete causal wedge and its boundary diamond map to a partial accelerated-mirror problem in Minkowski space. Second, we solve that flat problem directly in polar Rindler coordinates and exhibit the exponentially growing modes responsible for the failure of an ordinary smearing kernel. Finally, we transform the problem to the common \(\mathbb R\times H^2\) frame, where the same growth reproduces the standard AdS–Rindler result.
 
\subsection{The causal wedge as a partial mirror problem}
\label{subsec:wedge-flat-image}

The purpose of this subsection is to establish that the entire AdS causal
wedge and its complete boundary diamond map into the flat partial-mirror
problem, so that no part of the reconstruction region or its boundary data is
lost under the conformal map.

Consider the boundary half-circle

 \begin{equation}
 A=
 \left\{
 \tau=0,\quad
 \rho=\frac{\pi}{2},\quad
 |\theta|<\frac{\pi}{2}
 \right\}.
 \end{equation}

Its boundary domain of dependence is

 \begin{equation}
 D[A]
 =
 \left\{
 \rho=\frac{\pi}{2},\quad
 |\tau|+|\theta|<\frac{\pi}{2}
 \right\},
 \end{equation}

and the corresponding bulk causal wedge is

 \begin{equation}
 \sin\rho\cos\theta>|\sin\tau|.
 \label{eq:global-wedge}
 \end{equation}

Use the standard ESU-to-Minkowski map

 \begin{equation}
 D=\cos\tau+\cos\rho,
 \end{equation}

 \begin{equation}
 T=\frac{\sin\tau}{D},
 \qquad
 X=\frac{\sin\rho\cos\theta}{D},
 \qquad
 Y=\frac{\sin\rho\sin\theta}{D}.
 \label{eq:esu-minkowski-map}
 \end{equation}

Writing \(r^2=X^2+Y^2\), the wedge condition becomes

 \begin{equation}
 X>|T|,
 \end{equation}

while

 \begin{equation}
 r^2-T^2
 =
 \frac{\cos\tau-\cos\rho}
 {\cos\tau+\cos\rho}
 <1.
 \end{equation}

Hence the complete open causal wedge maps to

 \begin{equation}
 \boxed{
 W=
 \left\{
 X>|T|,
 \quad
 r^2-T^2<1
 \right\}.
 }
 \label{eq:flat-partial-wedge}
 \end{equation}

On the AdS boundary, \(\rho=\pi/2\), the same map reduces to

 \begin{equation}
 T=\tan\tau,
 \qquad
 X=\sec\tau\cos\theta,
 \qquad
 Y=\sec\tau\sin\theta,
 \end{equation}

and therefore

 \begin{equation}
 r^2-T^2=1.
 \end{equation}

The complete boundary diamond consequently maps to the mirror patch

 \begin{equation}
 \boxed{
 \Sigma_R=
 \left\{
 X>|T|,
 \quad
 r^2-T^2=1
 \right\}.
 }
 \label{eq:flat-mirror-patch}
 \end{equation}

Thus \(W\) is the portion of the mirror interior lying in the right Rindler
wedge, and \(\Sigma_R\) is its complete timelike boundary. Every point of the
open causal wedge lies inside the Minkowski conformal patch; only the tips of
the boundary diamond reach its conformal boundary. Reconstruction from
\(D[A]\) is therefore mapped to reconstruction from the full mirror patch
\(\Sigma_R\), without requiring any additional exterior or horizon data.
 \subsection{Smearing function divergence in polar Rindler coordinates}
\label{subsec:polar-rindler}

We now solve the conformal flat space analogue polar Rindler coordinates will help us see the divergence in the AdS-Rindler smearing function in an elementary way. The mapping between the two problems will be established in the next section.

Begin with ordinary Rindler coordinates,

 \begin{equation}
 T=\xi\sinh\chi,
 \qquad
 X=\xi\cosh\chi,
 \end{equation}

and introduce polar coordinates in the \((\xi,Y)\)-plane,

 \begin{equation}
 \xi=\varrho\cos\vartheta,
 \qquad
 Y=\varrho\sin\vartheta.
 \end{equation}

Equivalently,

 \begin{equation}
 T=\varrho\cos\vartheta\sinh\chi,
 \qquad
 X=\varrho\cos\vartheta\cosh\chi,
 \qquad
 Y=\varrho\sin\vartheta.
 \label{eq:polar-rindler-map}
 \end{equation}

The coordinate ranges

 \begin{equation}
 0<\varrho<1,
 \qquad
 -\frac{\pi}{2}<\vartheta<\frac{\pi}{2},
 \qquad
 \chi\in\mathbb R
 \end{equation}

cover the complete flat wedge. Indeed,

 \begin{equation}
 X>|T|
 \end{equation}

is equivalent to \(\cos\vartheta>0\), while

 \begin{equation}
 X^2+Y^2-T^2
 =
 \xi^2+Y^2
 =
 \varrho^2.
 \end{equation}

Consequently,

 \begin{equation}
 W
 =
 \left\{
 X>|T|,
 \quad
 X^2+Y^2-T^2<1
 \right\}
 \end{equation}

is simply \(0<\varrho<1\), and the accelerated mirror is the constant-radius
surface

 \begin{equation}
 \varrho=1.
 \end{equation}

In these coordinates the Minkowski metric is

 \begin{equation}
 ds_{\mathrm M}^2
 =
 -d\varrho^2
 +
 \varrho^2
 \left(
 \cos^2\vartheta\,d\chi^2-d\vartheta^2
 \right).
 \label{eq:polar-rindler-metric}
 \end{equation}

The expression in parentheses is the static-patch metric on unit
two-dimensional de Sitter space,

 \begin{equation}
 ds_{dS_2}^2
 =
 \cos^2\vartheta\,d\chi^2-d\vartheta^2.
 \end{equation}

Thus the flat wedge is a Lorentzian cone over a \(dS_2\) static patch, and the
mirror is its unit-radius section. The inward-pointing unit normal at the
mirror is

 \begin{equation}
 n=-\partial_\varrho,
 \end{equation}

and the induced boundary measure is

 \begin{equation}
 d\Sigma_{\mathrm M}
 =
 \cos\vartheta\,d\chi\,d\vartheta.
 \end{equation}

Since the flat scalar curvature vanishes, the conformally coupled equation is
simply

 \begin{equation}
 \Box_{\mathrm M}\Phi=0.
 \end{equation}

The flat wave operator takes the separated form

 \begin{equation}
 \Box_{\mathrm M}
 =
 -\partial_\varrho^2
 -\frac{2}{\varrho}\partial_\varrho
 +\frac{1}{\varrho^2}\Box_{dS_2},
 \label{eq:flat-cone-wave-operator}
 \end{equation}

where

 \begin{equation}
 \Box_{dS_2}
 =
 \sec^2\vartheta\,\partial_\chi^2
 -
 \frac{1}{\cos\vartheta}
 \partial_\vartheta
 \left(
 \cos\vartheta\,\partial_\vartheta
 \right).
 \end{equation}

Let \(Y_\lambda(\chi,\vartheta)\) be any tangential mode satisfying

 \begin{equation}
 \Box_{dS_2}Y_\lambda
 =
 \lambda Y_\lambda,
 \end{equation}

and write

 \begin{equation}
 \Phi(\varrho,\chi,\vartheta)
 =
 R_\lambda(\varrho)Y_\lambda(\chi,\vartheta).
 \end{equation}

The flat wave equation then reduces to the Euler equation

 \begin{equation}
 R_\lambda''
 +\frac{2}{\varrho}R_\lambda'
 -\frac{\lambda}{\varrho^2}R_\lambda
 =0.
 \label{eq:flat-radial-euler}
 \end{equation}

Define

 \begin{equation}
 \kappa=\sqrt{\lambda+\frac14}.
 \end{equation}

The two radial solutions are

 \begin{equation}
 \varrho^{-1/2+\kappa},
 \qquad
 \varrho^{-1/2-\kappa}.
 \end{equation}

For reconstruction from the physical flat-space normal response, the radial
mode must satisfy

 \begin{equation}
 R_\lambda(1)=0,
 \qquad
 -R_\lambda'(1)=1.
 \label{eq:flat-response-normalization}
 \end{equation}

These conditions give

 \begin{equation}
 \boxed{
 R_\lambda^D(\varrho)
 =
 \frac{
 \varrho^{-1/2-\kappa}
 -
 \varrho^{-1/2+\kappa}
 }{2\kappa}.
 }
 \label{eq:flat-response-mode}
 \end{equation}

No horizon condition has been imposed: the homogeneous Dirichlet condition
and the prescribed normal response already fix both radial coefficients.

The origin of the instability is now immediate. For every fixed interior
point \(0<\varrho<1\),

 \begin{equation}
 R_\lambda^D(\varrho)
 \sim
 \frac{1}{2\kappa}
 \varrho^{-1/2-\kappa}
 =
 \frac{\varrho^{-1/2}}{2\kappa}
 \exp\!\left[
 \kappa\log\!\left(\frac1\varrho\right)
 \right]
 \end{equation}

as \(\kappa\rightarrow+\infty\). High positive tangential eigenvalues are
therefore amplified exponentially as the response is continued away from the
timelike mirror.

This can be written in an even more familiar form. Set

 \begin{equation}
 u=-\log\varrho,
 \qquad
 \Psi=\varrho^{1/2}\Phi.
 \end{equation}

Then

 \begin{equation}
 ds_{\mathrm M}^2
 =
 \varrho^2
 \left(
 ds_{dS_2}^2-du^2
 \right),
 \end{equation}

and the wave equation becomes

 \begin{equation}
 \left(
 \Box_{dS_2}
 -\partial_u^2
 +\frac14
 \right)\Psi=0.
 \end{equation}

For a mode with

 \begin{equation}
 \Box_{dS_2}Y_\lambda
 =
 \left(\kappa^2-\frac14\right)Y_\lambda,
 \end{equation}

the radial equation is simply

 \begin{equation}
 \frac{d^2\Psi_\kappa}{du^2}
 =
 \kappa^2\Psi_\kappa.
 \end{equation}

The response-normalized solution is

 \begin{equation}
 \boxed{
 \Psi_\kappa(u)
 =
 \frac{\sinh(\kappa u)}{\kappa}.
 }
 \label{eq:flat-sinh-amplification}
 \end{equation}

Thus the obstruction is the standard \(\sinh(\kappa u)\) amplification
encountered when high-spatial-momentum data are continued away from a
timelike boundary.  

A completely elementary family already exhibits the divergence. Consider
the \(\chi\)-independent sector and set

 \begin{equation}
 x=\sin\vartheta.
 \end{equation}

The Legendre polynomials satisfy

 \begin{equation}
 \Box_{dS_2}P_n(x)
 =
 n(n+1)P_n(x).
 \end{equation}

For these modes,

 \begin{equation}
 \kappa=n+\frac12,
 \end{equation}

and the exact flat-space solutions are

 \begin{equation}
 \boxed{
 \Phi_n(\varrho,\vartheta)
 =
 \frac{
 \varrho^{-n-1}-\varrho^n
 }{2n+1}
 P_n(\sin\vartheta).
 }
 \label{eq:elementary-flat-modes}
 \end{equation}

They obey

 \begin{equation}
 \Phi_n(1,\vartheta)=0,
 \end{equation}

and their physical mirror responses are

 \begin{equation}
 -\left.
 \partial_\varrho\Phi_n
 \right|_{\varrho=1}
 =
 P_n(\sin\vartheta).
 \end{equation}

These responses remain uniformly bounded,

 \begin{equation}
 \left|P_n(x)\right|\leq1,
 \qquad
 -1\leq x\leq1,
 \end{equation}

whereas the corresponding bulk fields grow exponentially with \(n\). For
example,

 \begin{equation}
 P_{2m}(0)
 =
 (-1)^m
 \frac{(2m)!}{2^{2m}(m!)^2}
 \sim
 \frac{(-1)^m}{\sqrt{\pi m}},
 \end{equation}

so at \(\vartheta=0\),

 \begin{equation}
 \left|
 \Phi_{2m}(\varrho,0)
 \right|
 \sim
 \frac{
 \varrho^{-2m-1}
 }{
 4m\sqrt{\pi m}
 }
 \end{equation}

for every fixed \(0<\varrho<1\). Thus a bounded sequence of increasingly
oscillatory mirror responses produces exponentially large values at an
interior point.

The same conclusion appears directly in the static smearing series. Since

 \begin{equation}
 \int_{-1}^{1}
 P_n(x)P_m(x)\,dx
 =
 \frac{2}{2n+1}\delta_{nm},
 \end{equation}

the static response kernel would be

 \begin{align}
 K_{\mathrm{static}}(\varrho;x,x')
 &=
 \sum_{n=0}^{\infty}
 \frac{2n+1}{2}
 R_n^D(\varrho)
 P_n(x)P_n(x')
 &=
 \frac12
 \sum_{n=0}^{\infty}
 \left(
 \varrho^{-n-1}-\varrho^n
 \right)
 P_n(x)P_n(x').
\label{eq:static-flat-kernel}
 \end{align}

For \(0<\varrho<1\), the second term is convergent, but the first contains
powers of \(1/\varrho>1\). At \(x=x'=0\), its even terms behave as

 \begin{equation}
 \frac12
 \varrho^{-2m-1}
 P_{2m}(0)^2
 \sim
 \frac{
 \varrho^{-2m-1}
 }{
 2\pi m
 },
 \end{equation}

and therefore do not even tend to zero. The static kernel already diverges;
including time-dependent modes cannot restore it.

 \subsection{Conformal Map of AdS-Rindler HKLL}
\label{subsec:H2-cross-check}

The polar-Rindler calculation above establishes the obstruction directly in
flat space. We now briefly relate it to the standard AdS--Rindler description.
The purpose of this subsection to
show that the flat partial-mirror problem and the AdS--Rindler problem are
conformal descriptions of the same boundary reconstruction problem, and to
recover the familiar large-transverse-momentum rate in AdS--Rindler
coordinates.

In ordinary flat Rindler coordinates,

 \begin{equation}
 T=\xi\sinh\chi,
 \qquad
 X=\xi\cosh\chi,
 \end{equation}

the metric is

 \begin{equation}
 ds_{\mathrm M}^2
 =
 \xi^2d\chi^2-d\xi^2-dY^2,
 \end{equation}

and the accelerated mirror is the semicircle

 \begin{equation}
 \xi^2+Y^2=1,
 \qquad
 \xi>0.
 \end{equation}

Introduce Fermi coordinates \((\sigma,q)\) about this semicircle by

 \begin{equation}
 \xi
 =
 \frac{1}
 {\cosh\sigma\cosh q+\sinh\sigma},
\qquad
 Y
 =
 \frac{\cosh\sigma\sinh q}
 {\cosh\sigma\cosh q+\sinh\sigma}.
 \label{eq:H2-fermi-map}
 \end{equation}

They satisfy

 \begin{equation}
 \frac{d\xi^2+dY^2}{\xi^2}
 =
 d\sigma^2+\cosh^2\sigma\,dq^2.
 \end{equation}

Consequently,

 \begin{equation}
 ds_{\mathrm M}^2
 =
 \xi^2ds_0^2,
\qquad
 ds_0^2
 =
 d\chi^2-d\sigma^2-\cosh^2\sigma\,dq^2.
 \label{eq:H2-common-metric}
 \end{equation}

The mirror is the geodesic \(\sigma=0\), and its interior is
\(\sigma>0\). The common metric \(ds_0^2\) is the product metric on
\(\mathbb R\times H^2\).

The AdS--Rindler metric takes the same form. Setting

 \begin{equation}
 z=\tanh\sigma
 \end{equation}

in

 \begin{equation}
 ds_{\mathrm{AdS}}^2
 =
 \frac{1}{z^2}
 \left[
 (1-z^2)d\chi^2
 -\frac{dz^2}{1-z^2}
 -dq^2
 \right]
 \end{equation}

gives

 \begin{equation}
 ds_{\mathrm{AdS}}^2
 =
 \frac{1-z^2}{z^2}ds_0^2
 =
 \frac{1}{\sinh^2\sigma}ds_0^2.
 \label{eq:adsrindlerH2}
 \end{equation}

Thus

 \begin{equation}
 ds_{\mathrm M}^2=\xi^2ds_0^2,
\qquad
 ds_{\mathrm{AdS}}^2
 =
 \sinh^{-2}\sigma\,ds_0^2.
 \end{equation}

This is the conformal relation between the flat partial-mirror wedge
and the AdS--Rindler wedge.

For a conformally coupled scalar in three dimensions, a Weyl transformation

 \begin{equation}
 g'_{\mu\nu}=\Omega^2g_{\mu\nu}
 \end{equation}

acts on the field as

 \begin{equation}
 \Phi'=\Omega^{-1/2}\Phi.
 \end{equation}

Therefore, if \(\Phi_0\) denotes the field in the common frame,

 \begin{equation}
 \Phi_{\mathrm M}=\xi^{-1/2}\Phi_0,
\qquad
 \Phi_{\mathrm{AdS}}
 =
 (\sinh\sigma)^{1/2}\Phi_0.
 \label{eq:H2-field-transport}
 \end{equation}

Let

 \begin{equation}
 \mathcal F_0(\chi,q)
 =
 \left.
 \partial_\sigma\Phi_0
 \right|_{\sigma=0}
 \end{equation}

be the common-frame normal response. On the flat mirror,

 \begin{equation}
 \xi_b=\operatorname{sech}q,
 \end{equation}

and the physical flat response and measure are

 \begin{equation}
 \mathcal F_{\mathrm M}
 =
 \xi_b^{-3/2}\mathcal F_0,
\qquad
 d\Sigma_{\mathrm M}
 =
 \xi_b^2\,d\chi\,dq.
 \end{equation}

Hence, if \(K_0\) is the common-frame kernel, the physical flat kernel is

 \begin{equation}
 K_{\mathrm M}(P|b)
 =
 \xi_P^{-1/2}\xi_b^{-1/2}K_0(P|b).
 \label{eq:H2-flat-kernel-transport}
 \end{equation}

Near the AdS boundary,

 \begin{equation}
 \Phi_0
 =
 \sigma\,\mathcal F_0+O(\sigma^2),
 \end{equation}

so

 \begin{equation}
 \Phi_{\mathrm{AdS}}
 =
 z^{3/2}\mathcal F_0+O(z^{5/2}).
 \end{equation}

Thus \(\mathcal F_0\) is also the normalizable AdS boundary coefficient. The
common-frame kernel therefore transports consistently to both the physical
flat response and the AdS boundary response.

With the curvature conventions used here, the field equation in the common
frame is

 \begin{equation}
 \left(
 \Box_0-\frac14
 \right)\Phi_0=0.
 \end{equation}

Fourier decomposition along the boundary geodesic,

 \begin{equation}
 \Phi_0
 =
 e^{-i\omega\chi+ikq}
 g_{\omega k}(\sigma),
 \end{equation}

gives

 \begin{equation}
 g_{\omega k}''
 +\tanh\sigma\,g_{\omega k}'
 +
 \left[
 \omega^2+\frac14
 -k^2\operatorname{sech}^2\sigma
 \right]g_{\omega k}
 =0.
 \label{eq:fermiODE}
 \end{equation}

The response-normalized Dirichlet mode is fixed by

 \begin{equation}
 g_{\omega k}(0)=0,
 \qquad
 g_{\omega k}'(0)=1.
 \end{equation}

These two boundary conditions determine the mode uniquely; no condition at
the Rindler horizon is required. The exact solution can be expressed in
hypergeometric functions, but its detailed form is not needed here.

Writing

 \begin{equation}
 g_{\omega k}
 =
 (\cosh\sigma)^{-1/2}f_{\omega k}
 \end{equation}

reduces the equation to

 \begin{equation}
 f_{\omega k}''
 +
 \left[
 \omega^2
 -
 \left(k^2+\frac14\right)
 \operatorname{sech}^2\sigma
 \right]f_{\omega k}
 =0.
 \end{equation}

For fixed \(\sigma>0\), bounded \(\omega\), and
\(|k|\rightarrow\infty\),

 \begin{equation}
 \log
 \left|
 g_{\omega k}(\sigma)
 \right|
 =
 |k|
 \int_0^\sigma
 \operatorname{sech}s\,ds
 +O(\log|k|).
 \end{equation}

Since

 \begin{equation}
 \int_0^\sigma
 \operatorname{sech}s\,ds
 =
 \arcsin(\tanh\sigma),
 \end{equation}

one obtains

 \begin{equation}
 \boxed{
 \log
 \left|
 g_{\omega k}(\sigma)
 \right|
 =
 |k|\arcsin(\tanh\sigma)
 +O(\log|k|).
 }
 \label{eq:gdrate}
 \end{equation}

The formal common-frame kernel,

 \begin{equation}
 K_0(\sigma;\Delta\chi,\Delta q)
 =
 \int
 \frac{d\omega\,dk}{(2\pi)^2}
 e^{-i\omega\Delta\chi+ik\Delta q}
 g_{\omega k}(\sigma),
 \label{eq:rindlerkernel}
 \end{equation}

therefore fails to define an ordinary function at every interior point. The
local Weyl factors cannot remove this
exponential large-\(|k|\) growth.

This is the same instability found directly in polar Rindler coordinates,
where the response-normalized radial factor was

 \begin{equation}
 \varrho^{-1/2}
 \frac{
 \sinh\!\left[
 \kappa\log(1/\varrho)
 \right]
 }{\kappa}.
 \end{equation}

The parameters \(k\) and \(\kappa\) belong to different tangential mode
decompositions and should not be identified mode by mode. What is invariant
is the underlying mechanism: sufficiently rapid variation along the
timelike mirror excites an exponentially growing branch when the response is
continued into the bulk. The \(\mathbb R\times H^2\) frame reproduces the
standard AdS--Rindler expression for this growth, while the polar-Rindler
frame exposes its elementary flat-space origin.

\section{Conclusions}

In the paper, we conformally mapped the problem of HKLL reconstruction in AdS$_3$ for a conformally coupled scalar to two different problems: A Dirichlet problem in ESU and a mirror problem with restricted data in flat space. These maps are instructive as they provide different perspectives on a known problem. 

A particular point of interest was that both for ESU and flat space, the smearing function was not a solution of the equation of motion. However it satisfied the EoM up to the known ambiguities in the smearing function, which is the actual condition that a smearing function needs to satisfy. 

We also considered the case of causal wedge reconstruction. This was the case where the conformal map was perhaps most insightful. The problem of causal wedge reconstruction gets mapped to a partial mirror problem in flat space where the mode solutions have simple power law behavior. The well-known divergence of the smearing function in causal wedges has a simple manifestation in this setting. 

It would be natural to ask what the alternate quantization \(\Delta=1/2\) gives
(presumably a Neumann mirror condition), whether the same cap-data structure
arises for AdS--Rindler or for higher-dimensional and massive fields, and, at
the quantum level, which state of the mirror field the global AdS vacuum
corresponds to. We leave these for future investigation.
\begin{acknowledgments}
AI was used for language editing.   
\end{acknowledgments}
 \bibliographystyle{unsrt}
 \bibliography{ads3}

\end{document}